\documentclass[iop]{emulateapj}

\usepackage{ulem}
\usepackage{color}

\shorttitle{SS433's jet trace from ALMA imaging and Global Jet Watch spectroscopy}
\shortauthors{Blundell, Laing, Lee \& Richards}
\slugcomment{Accepted by ApJ Letters}

\begin{document}

\title{SS433's jet trace from ALMA imaging and Global Jet Watch spectroscopy:\\ evidence for post-launch particle acceleration}

\author{Katherine M.\ Blundell\altaffilmark{1} , Robert Laing\altaffilmark{2}, Steven Lee\altaffilmark{3} and Anita Richards\altaffilmark{4}, }

\altaffiltext{1}{University of Oxford, Department of Physics, Keble
  Road, Oxford, OX1 3RH, UK}
\altaffiltext{2}{SKA Organisation, Jodrell Bank Observatory, Lower Withington, Macclesfield, SK11 9DL, UK}
\altaffiltext{3}{Anglo-Australian Telescope, Coonabarabran, NSW 2357, Australia}
\altaffiltext{4}{Jodrell Bank Centre for Astrophysics, School of Physics and Astronomy, University of Manchester, Manchester, M13 9PL, UK}

\begin{abstract}
We present a comparison of Doppler-shifted H$\alpha$ line emission observed by the Global Jet Watch from freshly-launched jet ejecta at the nucleus of the Galactic microquasar SS433 with subsequent ALMA imaging at mm-wavelengths of {\it the same} jet ejecta.  There is a remarkable similarity between the transversely-resolved synchrotron emission and the prediction of the jet trace from optical spectroscopy:  this is an a priori prediction not an a posteriori fit, confirming the ballistic nature of the jet propagation.  
The mm-wavelength of the ALMA polarimetry is sufficiently short that the Faraday rotation is negligible and therefore that the observed ${\bf E}$-vector directions are accurately orthogonal to the projected local magnetic field.  
Close to the nucleus the ${\bf B}$-field vectors are perpendicular to the direction of propagation.  Further out from the nucleus, the ${\bf B}$-field vectors that are coincident with the jet instead become parallel to the ridge line; this occurs at a distance where the jet bolides are expected to expand into one another.  X-ray variability has also been observed at this location; this has a natural explanation if shocks from the expanding and colliding bolides cause particle acceleration.  In regions distinctly separate from the jet ridge line, the fractional polarisation approaches the theoretical maximum for synchrotron emission.  
\end{abstract}

\section{Introduction}
Since shortly after its discovery four decades ago the prototypical Galactic microquasar SS433 has been known to eject oppositely-directed jets whose launch axis precesses with a cone angle of about 19 degrees approximately every 162 days and whose speeds average to about a quarter of the speed of light.  Striking images of the emission at radio (cm) wavelengths reveal a zigzag/corkscrew structure that arises because of the above properties modulated by light-travel time effects arising from its orientation with respect to our line-of-sight \citep{Hjellming1981,Stirling2002,Blundell04,Roberts2008,Miller-Jones2008}.  The optical spectra of this object are characterised by a strong Balmer H$\alpha$ emission line complex close to the rest-wavelength of this line, and also blue-shifted and red-shifted lines whose observed wavelengths change successively on a daily basis according to the instantaneous speed and angle of travel with respect to our line of sight.  Fitted parameters to the kinematic model developed from the first few years of optical spectroscopy were presented by e.g.\ \citet{Margon84} and \citet{Eikenberry01}.   Hitherto the timing of optical spectroscopy and spatially resolved radio imaging has not permitted the observation of the same ejecta both at launch and after propagation.  
   We present the first mm-wave image of SS433 from ALMA in combination with optical spectroscopy (Sec\,\ref{sec:SpeedsFromZeds}) of the same ejecta observed during the year prior to the ALMA observations (Sec\,\ref{sec:ALMA}).  This allows us to distinguish ballistic motion post-launch from deceleration \citep[e.g.][]{Stirling2004}.

A long standing question is why SS433's jet ejecta are primarily line emitting at launch yet synchrotron emitting at largest distances from the nucleus; the polarisation changes explored in Sec.\,\ref{sec:polar} shed light on this.   Inference of the magnetic-field structure in the jets is complicated by the combined effects of Faraday rotation and time-variable
structure. Previous studies \citep{Stirling2004,Roberts2008,Miller-Jones2008} have been hampered
by lack of resolution and frequency coverage as well as the uncertain
effects of spatial- and temporal-variations in Faraday rotation.  The dependence of Faraday rotation on the square of the wavelength ($\lambda$), means that wide-band observations at mm wavelengths allow the projected field direction to be determined accurately in a single observation even close to the core, where Faraday rotation measures may be large \citep{Roberts2008}.  We present our polarimetric 230\,GHz results in Sec.\,\ref{sec:polar}.

\section{Optical spectroscopy and inference from Doppler shifts}
\label{sec:SpeedsFromZeds}
Spectra of SS433 spanning a wavelength range of approximately 5800 to 8500 Angstroms, and having a spectral resolution of $\sim4000$ were observed in the year prior to the ALMA observations whenever this target was a nighttime object.  These were carried out with the multi-longitude Global Jet Watch telescopes each of which is equipped with an Aquila spectrograph; the design and testing of these high-throughput spectrographs are described by \citet{Lee2018}.  The
observatories, astronomical operations, processing and calibration of the spectroscopic data streams are described in \citet{Blundell2018}. Almost all of these spectra contain a pair of so-called ``moving lines'' arising from the most recently launched jet bolides in SS433.  The wavelengths corresponding to the centroids of the blue-shifted and red-shifted H$\alpha$ emission were converted into redshift pairs with respect to H$\alpha$ in the rest frame of SS433 according to its systemic velocity with respect to Earth \citep{lockman2007}.  From these redshift and blueshift pairs from a given spectrum were derived the
launch speed of each pair of bolides \citep[equation 2]{Blundell05}. This avoids the approximation of constant ejection speed, which has been shown to be inaccurate from archival spectroscopy \citep{Blundell05,Blundell11}.  Assuming that the subsequent motion is ballistic (this assumption is discussed in Sec\,\ref{sec:ballistic}), and adopting the standard kinematic model \citep{Hjellming1981}, the locations they attain by the Julian Date of the mid-point of the ALMA observations (2457294.4836) are
calculated, and plotted in Fig\,\ref{fig:combine}.  The assumed
parameters of the kinematic model using the notation
of \citet{Eikenberry01} and \citet{Hjellming1981} are: cone angle $\theta = 19^\circ$ (Hjellming et al. use $\psi$), inclination $i = 79^\circ$, rotation on the sky $\chi = 10^\circ$ (position angle $+100^\circ$), period $P = 162.34$\,day (Blundell et al., in preparation) and distance $d =
5.5$\,kpc \citep{Blundell04,lockman2007}. The ejection phase was determined by fitting to the observed redshift pairs from JD 2457000 to JD 2457293.5. The phase
$\phi = (2\pi/P)(t - t_{\rm ref}) + \phi_0$ with $\phi_0 =
-0.241$\,rad for a reference Julian date of $t_{\rm ref} =
2456000$. $\phi$ is used as in equation 1 of \citet{Eikenberry01}; \citet{Hjellming1981} denote the same
quantity by $\Omega(t_0-t_{\rm ref})$.

\section{Millimetre polarimetric imaging}
\label{sec:ALMA}

SS433 was observed using 27 ALMA antennas between 2015 September 28 
21:26 and  September 29 01:46 UT. Three execution blocks were run almost in sequence and under 
similar conditions. The
precipitable water vapour column was around 1.4\,mm.  The correlator was set up in Time Division Multiplex mode with 
a total bandwidth of 7.5\,GHz, in four 1.75-GHz spectral windows (spw) 
centred at 224, 226, 240 and 242\,GHz. Each spw was divided into 64 spectral channels 
and XX, YY, XY and YX correlations were recorded.  The longest and
shortest baselines were 2270 and 43\,m, sensitive to angular scales
$\la 3.7$\,arcsec.

The quasar J1751+0939 was used as a bandpass, polarization and flux
scale calibrator and J1832+0731 was used as the phase reference source
on an approximately 8 min cycle. The total integration time on SS433
was $\approx$2\,hr.  Initial data reduction followed standard ALMA
scripts, executed in CASA \citep{Schnee2014}.  The flux density of
J1751+0939 during these observations was taken to be 3.7275\,Jy at
232.86\,GHz with a spectral index $\alpha = -$0.441 (defined in the sense $S(\nu) \propto \nu^{-\alpha}$) and the total flux scale
uncertainty is about 10\%.  Polarization leakage was calibrated as
described by \citet{Nagai16}.
Several iterations of {\sc clean} in multi-frequency synthesis mode \citep{Rau} 
followed by self-calibration were used to improve the imaging of
SS433. The final iteration of amplitude and phase self-calibration was
made by combining all four spectral windows using a model with two
Taylor series terms.  We show the zero-order Taylor series $I$ image
after self-calibration, together with polarised intensity images derived from $Q$ and $U$ for the
entire band (we demonstrate below that Faraday rotation is negligible
for our frequency range).  The off-source rms levels are 13, 11 and
12\,$\mu$Jy\,beam$^{-1}$ in $I$, $Q$ and $U$, respectively, consistent
with the expectations for thermal noise alone. The restoring beam has FWHM 
$0.19 \times 0.16$\,arcsec$^2$. 

The $I$ image (Fig\,\ref{fig:combine}, central panel greyscale) shows the familiar zigzag/corkscrew shape of SS433. The peak flux density at 233\,GHz is 86.0 mJy/beam.  The in-band spectral index of the core is $-0.29 \pm 0.14$.

\section{Comparison of time-extrapolated spectroscopy with ALMA imaging}
\label{sec:ballistic}

\begin{figure}[htbp]
\begin{center}
    \hspace*{-1cm}
    \includegraphics[trim=0cm 0cm 0cm 0cm, clip,width=0.53\textwidth]{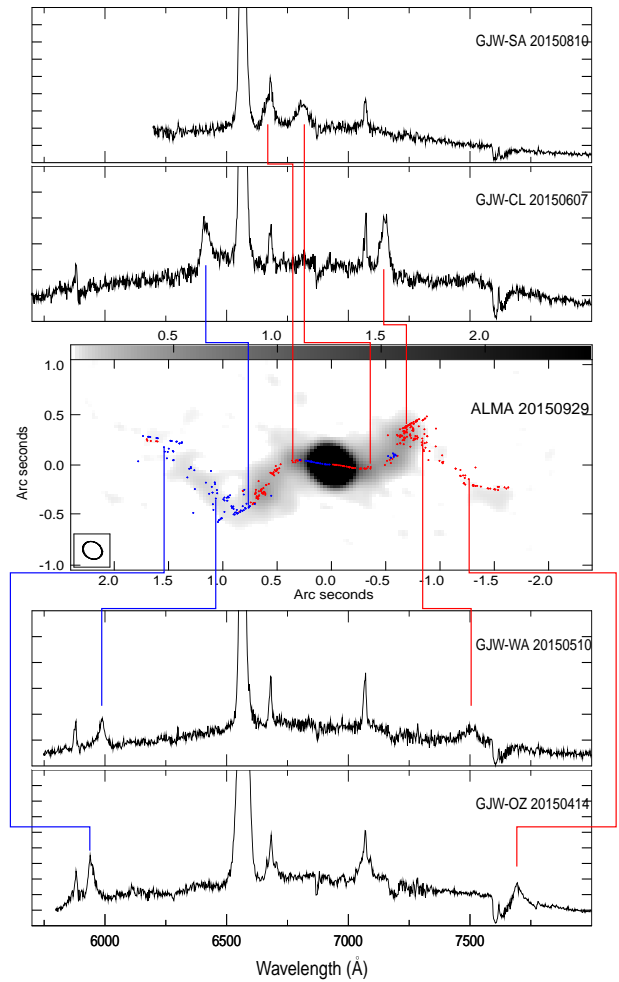} 
\caption{{\it Central panel:}  mm-wave image made from ALMA observations in 2015 September described in Sec\,\ref{sec:ALMA} overlaid with symbols representing the positions attained by the bolides whose speeds are measured via optical spectroscopy and assumed to move ballistically after launch.  {\it Upper two and lower two panels:} Representative spectra from four different dates prior to the ALMA observations are shown, revealing a pair of Doppler shifted lines corresponding to emission from the oppositely moving jet bolides, recently launched and still optically radiant.  These spectra are from each of four different observatories, from bottom to top, namely eastern Australia (GJW-OZ), Western Australia (GJW-WA), Chile (GJW-CL) and South Africa (GJW-SA).   }
 \label{fig:combine}
 \end{center}
 \end{figure}

Fig\,\ref{fig:combine} shows excellent agreement between the jet trace predicted by the redshift and blueshift pairs from the optical spectroscopy and the brightness distribution subsequently measured by ALMA at mm-wavelengths.  There are only three free parameters in the superposition: two positional offsets to align the ejection centre with the peak of the radio emission and the rotation of the precession axis on the sky (which was taken from \citet{Hjellming1981}, not determined independently).  This is the first time that it has been possible to demonstrate the superposition directly with optical spectroscopy covering the appropriate time period. We find no evidence for significant deceleration of the jet post-launch (which would be evinced by radio emission systematically lagging the optical bolides).  In particular, we can rule out a deceleration of 0.02$c$ just outside the launch location as suggested\footnote{The lower speeds suggested by \citet{Stirling2004}  are a direct consequence of their adoption of a smaller distance $d = 4.8$\,kpc for SS433.} by \citet{Stirling2004}: this would lead to a systematic offset of $\approx$0.3\,arcsec between the predicted jet trace and the ALMA brightness distribution at projected distances $\ga$1\,arcsec, most obviously on the East side of the source; this is not seen.  More stringent constraints on deceleration can be obtained from a comparison of the jet trace predicted by optical spectroscopy with VLA radio imaging between 8\,GHz and 12\,GHz, in which the trace is detectable out to much greater distances from the nucleus than is possible in our current 230-GHz observations. We will address this comparison in a future paper, together with possible correlations of speed with launch angle.

\section{$B$-field structure}
\label{sec:polar}

The apparently conflicting results in the literature for the
relation between the projected magnetic-field direction and the
underlying jet flow in SS433 can be understood by consideration of the different distances from the nucleus probed by these studies. \citet{Stirling2004} and \citet{Roberts2008}
found a preferential alignment between the magnetic field and the jet
ridge line from $\approx$0.4 -- 2\,arcsec from the nucleus, whereas figure 8 of \citet{Miller-Jones2008} suggests that the field is
instead parallel to the ballistic velocity of the jet knots at
distances larger than 2\,arcsec. 

\begin{figure}[htbp]
\begin{center}
   \includegraphics[width=6.5cm]{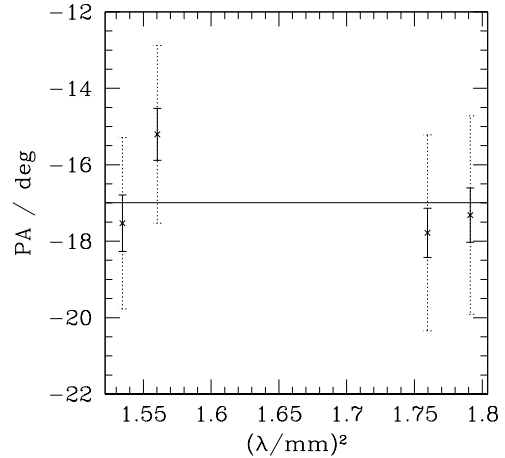}
\caption{The direction of the ${\bf B}$-field at four different wavelengths plotted as a function of $\lambda^2$.  All measurements are consistent with $-17$ degrees.  The solid bars represent errors due to thermal noise alone while the dotted bars include systematic errors. 
\label{fig:paplot}}
\end{center}
\end{figure}

\begin{figure}[htbp]
\begin{center}
   \includegraphics[width=8.5cm]{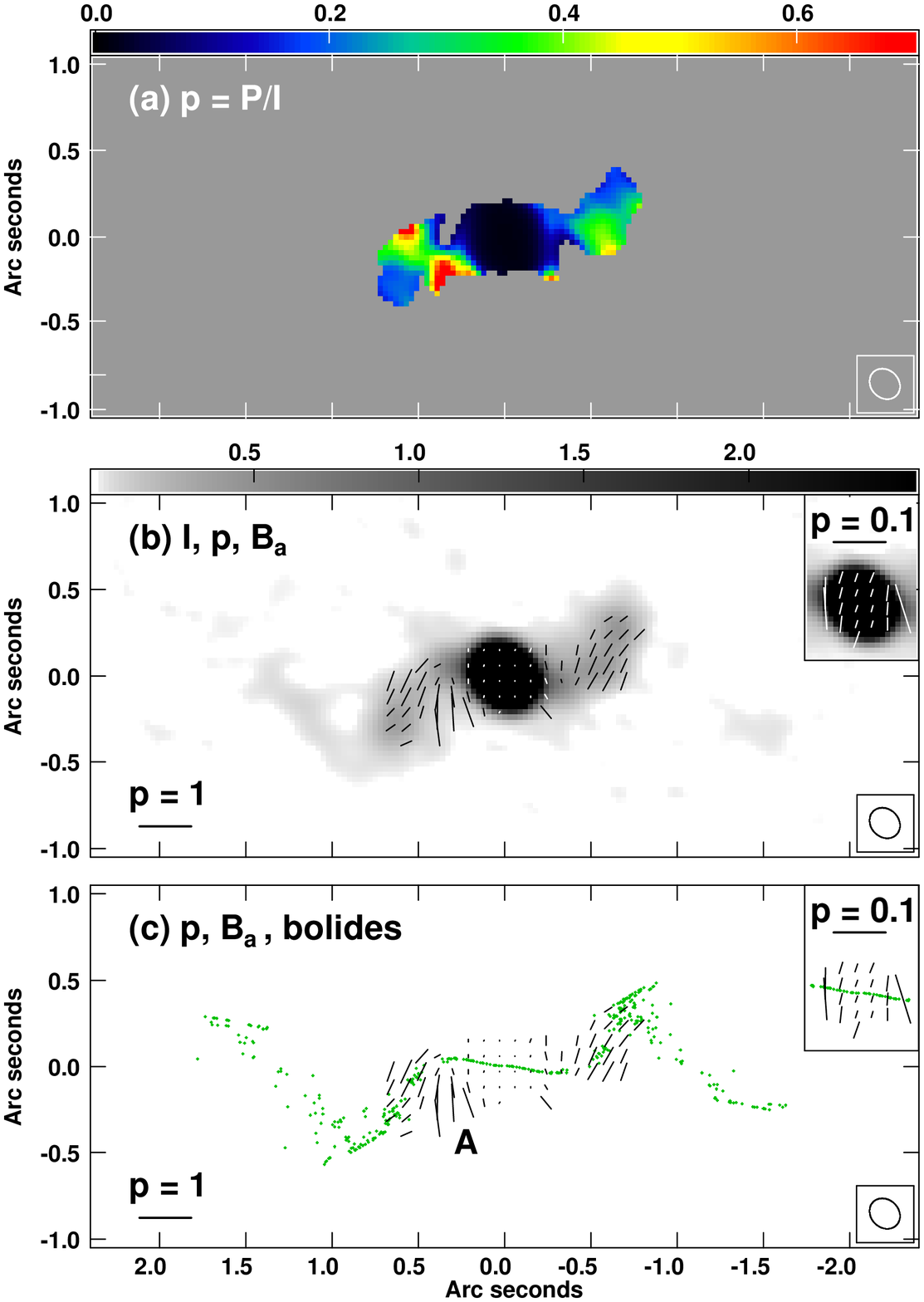} 
\caption{Panel (a) shows a colour scale depicting the fractional polarisation, $p$, of SS433 averaged over the ALMA observing band. Points are shown blanked (grey) wherever the total intensity $I < 5\sigma_I$. Panel (b) shows vectors whose lengths are proportional to $p$ (with the scale indicated by the labelled bar) and whose directions are along the apparent ${\bf B}$-field direction (i.e. rotated by 90$^\circ$ from the ${\bf E}$-vector direction with no correction for Faraday rotation: see text). The vectors are plotted where $I > 5\sigma_I$ and $P > 3\sigma_I$ and are superimposed on a grey-scale of total intensity, as indicated by the wedge labelled in mJy/beam. The inset shows the core with vectors plotted on an expanded scale for $I > 1$\,mJy/beam and $P > 3\sigma_I$.  Panel (c): as (b), but with vectors superimposed on the inferred locations reached by pairs of plasma bolides (green crosses, as in Fig\,\ref{fig:combine}).  The inset again shows the core region.}
 \label{fig:fracpolvectors}
 \end{center}
 \end{figure}

Our measurements of the magnetic-field direction are much less
affected by Faraday rotation than those in earlier
work. \citet{Stirling2004} found a mean rotation measure (RM) of
119\,rad\,m$^{-2}$ (excluding the core), which implies a position
angle rotation of 0.01$^\circ$ at 230\,GHz.  Even for the RM's of
$\approx$600\,rad\,m$^{-2}$ estimated for distances within 0.4\,arcsec
of the core \citep{Stirling2004,Roberts2008}, the inferred rotation is
still only 0.06$^\circ$ at this ALMA band.  We also find no evidence for any
wavelength-dependent rotation across our observing band. In
particular, the position angles measured for the individual spws at
the location of the core (where we might expect maximum Faraday
rotation) are consistent with the mean value for the band with an rms
scatter of 1.0$^\circ$ and Fig.~\ref{fig:paplot} shows no systematic trend. We therefore conclude that the
position angles plotted in Fig.~\ref{fig:fracpolvectors} are not
significantly affected by Faraday rotation.

While the fractional polarisation of the nucleus of SS433 at 230\,GHz
is low ($p = 0.011$), the position angle is still securely determined.
The inset to Fig\,\ref{fig:fracpolvectors} shows that within
$\approx$0.35\,arcsec of the nucleus the orientation of the $B$-field
vectors is consistent with being perpendicular to the jet ridge line (and also to line of radial ejection, which is indistinguishable
from it at this distance), consistent with the tentative suggestion by
\citet{Roberts2008}.  At a distance of $\approx$0.35\,arcsec, the degree of polarisation increases to $p \approx 0.1$. 
Here, the field directions in both jets become parallel to the ridge line and
clearly inconsistent with the direction of ballistic motion. 
This
relative orientation persists out to at least 0.7\,arcsec, beyond which we
cannot measure accurate position angles.  Our result is consistent
with that of \citet{Stirling2004}, but with higher angular resolution
and lower uncertainties from Faraday rotation.  For distances from the launch-point exceeding $\approx$2\,arcsec, Miller-Jones et al (2008) report the magnetic field of the jet as being parallel to the local velocity vector (using the value for rotation measure reported by Stirling et al 2004).

The transition in field direction at 0.35\,arcsec ($3 \times
10^{14}$\,m) may bear on the oft-debated question of whether the
outflow in SS433 is best described as a succession of independent
bolides or a continuous jet. It is interesting to compare this
distance with the point at which the expanding bow shocks surrounding
neighbouring bolides first intersect.  If we assume that one bolide is
ejected per day at a speed of 0.26$c$, their radial separation is
$\approx$6.7 $\times 10^{12}$\,m. If the shock expansion speed is
comparable with the expansion rate of the radio knots measured with
VLBI ($\approx 0.015c$; \citealt{Jeffrey2016}), then the shock fronts
will indeed expand into each other and interact when the bolides have
travelled $\approx 3 \times 10^{14}$\,m from the nucleus, roughly
where the change in field direction occurs.    Bolides will coalesce to form larger structures which will then cease to interact with one another, when the paths of successive bolide conglomerates are too angularly divergent.  Thereafter, the magnetic field observed to be associated with the jet trace will no longer reflect the details of the bolides as launched but rather their interactions with the (magnetised) medium through which they flow.  This magnetic field behaviour appears to dominate for distances from the launch-point exceeding $\approx$2\,arcsec.

Very highly polarised emission is observed away from the jet trace on the East side of the source (labelled A in Fig~\ref{fig:fracpolvectors}).  Both \citet{Roberts2008} and \citet{Miller-Jones2008} have drawn attention to these off-ridgeline regions being significantly more polarised than the jet ridgeline itself in VLA data at 15\,GHz and 8\,GHz respectively.  Our 230\,GHz data show significantly higher fractional polarisation values (0.6 -- 0.7) in these regions.   Comparison of all these data suggest the degree and direction of polarisation may change with precession period and possibly also distance from the nucleus.  

The interaction of bolides into larger coalescences has been reasoned above to occur where there is a change in polarisation behaviour namely at approximately 0.35\,arcsec from the nucleus.  We note that this coincides with the region reported by \citet{Migliari2002,Migliari2005} to show distinct X-ray variability which we suggest arises from shocks formed by the coalescence.   Such a mechanism would naturally give rise to the stochastic nature of the X-ray variations reported by \citet{Migliari2002,Migliari2005}.

\section{Conclusions}

Over one precession period of the jets in the Galactic microquasar SS433 is shown to be traced out at mm-wavelengths in our ALMA imaging.  This shows remarkable concordance with the predicted trace of the same ejecta from Global Jet Watch spectroscopy at earlier epochs.   

At mm-wavelengths the Faraday Rotation towards SS433 is negligible.  By 0.35\,arcsec from the nucleus the ${\bf B}$-field direction has changed from being perpendicular to parallel to the jet ridge line.   This occurs where bolides are expected to have expanded into one another, and where X-ray variability has been reported, consistent with the onset of particle acceleration and the change from line-emission at launch to dominant synchrotron emission further out.

\acknowledgments 
This paper makes use of the following ALMA data: ADS/JAO.ALMA\#2013.1.01369.S. ALMA is a partnership of ESO (representing its member states), NSF (USA) and NINS (Japan), together with NRC (Canada), MOST and ASIAA (Taiwan), and KASI (Republic of Korea), in cooperation with the Republic of Chile. The Joint ALMA Observatory is operated by ESO, AUI/NRAO and NAOJ.   A great many organisations and individuals have contributed to the success of the Global Jet Watch observatories and these are listed on {\tt www.GlobalJetWatch.net} but we particularly thank the University of Oxford and the Australian Astronomical Observatory.

\end{document}